\documentclass[useAMS,usenatbib]{mn2e}

\usepackage[dvips]{graphicx}

\title[A High Signal to Noise Ratio Map of the Sunyaev-Zel'dovich increment at 1.1 mm wavelength in Abell 1835]{A High Signal to Noise Map of the Sunyaev-Zel'dovich increment at 1.1 mm wavelength in Abell 1835}
\author[Horner, et al.]{P. F. Horner$^{1}$\thanks{E-mail:
Piers.Horner@astro.cf.ac.uk (PFH); Philip.Mauskopf@astro.cf.ac.uk (PDM)}, P. D.
Mauskopf$^{1}$\footnotemark[1], J. Aguirre$^2$, J. J. Bock$^3$, E. Egami$^4$, \and
J. Glenn$^5$, S. R. Golwala$^6$, G. Laurent$^5$, H. T. Nguyen$^3$ \and 
and J. Sayers$^6$\\
$^{1}$ Cardiff School of Physics and Astronomy, Cardiff University, 1 The Parade, Cardiff, CF24 3AA, Wales\\
$^2$ Department of Physics and Astronomy, University of Pennsylvania, 209 South 33rd Street, Philadelphia, PA 19104\\
$^3$ Jet Propulsion Laboratory, California Institute of Technology, 4800 Oak Grove Drive, Pasadena, CA91109\\
$^4$ Department of Astronomy/Steward Observatory, 933 North Cherry Avenue, Rm. N204, Tucson, AZ 85721-0065\\
$^5$ Center for Astrophysics and Space Astronomy \& Department of Astrophysical and Planetary Sciences,University of Colorado, 389 UCB, Boulder, CO 80309\\
$^6$ Division of Physics, Mathematics, \& Astronomy, California Institute of Technology, Mail Code 59-33,Pasadena, CA 91125}
\begin{document}

\date{Draft 2010 May 25}

\pagerange{\pageref{firstpage}--\pageref{lastpage}} \pubyear{2010}

\maketitle

\label{firstpage}

\begin{abstract}
We present an analysis of an 8 arcminute diameter map of the area around the galaxy cluster Abell 
1835 from jiggle map observations at a wavelength of 1.1 mm using the Bolometric Camera 
(Bolocam) mounted on the Caltech Submillimeter Observatory (CSO). The data is well described 
by a model including an extended Sunyaev-Zel'dovich (SZ) signal from the cluster gas plus 
emission from two bright background submm galaxies magnified by the gravitational lensing of 
the cluster. The best-fit values for the central Compton value for the cluster and the fluxes 
of the two main point sources in the field: SMM J140104+0252, and SMM J14009+0252 are found to 
be $y_{0}=(4.34\pm0.52\pm0.69)\times10^{-4}$, 6.5$\pm{2.0}\pm0.7$ mJy and 11.3$\pm{1.9}\pm1.1$ 
mJy, where the first error represents the statistical measurement error and the second error 
represents the estimated systematic error in the result. This measurement assumes the 
presence of dust emission from the cluster's central cD galaxy of $1.8\pm0.5$~mJy, based on 
higher frequency observations of Abell 1835. 
The cluster image represents one of the highest-significance SZ detections of 
a cluster in the positive region of the thermal SZ spectrum to date. The inferred central 
intensity is compared to other SZ measurements of Abell 1835 and this collection of results 
is used to obtain values for $y_{0} = (3.60\pm0.24)\times10^{-4}$ and the cluster peculiar 
velocity $v_{z} = -226\pm275$~km/s.
\end{abstract}

\begin{keywords}
infrared: galaxies -- galaxies: clusters: general -- methods: data analysis
\end{keywords}

\section{Introduction}

The Sunyaev Zel'dovich (SZ) effect \citep{Sunyaev70} is the redistribution of 
energy in the Cosmic Microwave Background (CMB) spectrum due to interations 
between CMB photons and hot electrons along the line of sight between the 
surface of last scattering and an observer. The main source for the SZ effect is 
from the hot gas that exists in the intra-cluster medium (ICM) of massive galaxy 
clusters \citep{Birkinshaw99, Carlstrom02, Rephaeli06}. The SZ actually comprises two effects.
The thermal effect consists of a dimming or decrement in the apparent brightness of the 
CMB towards a galaxy cluster at low frequencies and a corresponding brightening or 
increment at high frequencies with the null crossover point at approximately 215 GHz. The 
kinematic effect has the spectral dependence of a standard temperature shift in the 
CMB which has the same sign at all frequencies. The two effects can therefore be 
distinguished from each other with measurements at multiple frequencies.

Measurements of the amplitude of the SZ thermal distortion towards a cluster can be 
combined with measurements of the X-ray emission to determine the angular diameter 
distance, $d_{A}$, whose value depends on cosmology \citep{Birkinshaw91}. Estimates of 
the Hubble constant have been made using this technique for a number of clusters 
(e.g. \cite{Jones95, Grainge96, Holzapfel97, Tsuboi98, Mauskopf00, Reese03, Battistelli03, 
Udomprasert04}) and can be used to constrain cosmological models. In addition, because 
the SZ surface brightness for a cluster with a given mass is almost independent of 
redshift, SZ surveys can give information about the evolution of the number counts of 
clusters vs. redshift. This depends strongly on the evolution of the so-called dark 
energy or cosmological constant and therefore SZ surveys have been identified as one of 
the key probes of the nature of dark energy (e.g. \cite{Diego02, Weller02, DeDeo05, 
Albrecht06, Bhattacharya07}). A number of dedicated SZ surveys are already producing 
results, in particular the Atacama Cosmology Telescope (ACT) and the South Polar Telescope 
(SPT) \citep{Plagge09, Hincks09, High10}.

Most of the SZ detections reported to date have been made at low frequencies corresponding 
to the SZ decrement. Follow-up photometric or spectroscopic measurements of known clusters 
at higher frequencies would significantly improve the precision in the measurement of the 
kinematic SZ effect as well as constraining possible contamination in the low frequency data. 
Measurements at high frequencies corresponding to the SZ increment suffer from confusion 
from emission from dusty galaxies, including background high redshift galaxies amplified by 
the gravitational lensing of the cluster as well as increased atmospheric contamination from 
ground-based telescopes. Accurate measurement of the SZ increment requires a combination 
of angular resolution sufficient to resolve and remove point sources combined with high 
sensitivity and control of systematics necessary to detect the more diffuse SZ signal.

This paper presents analysis of 
observations of the galaxy cluster Abell 1835 using the Bolometric Camera 
(Bolocam) mounted on the Caltech Submillimeter Observatory (CSO), situated 
on the summit of Mauna Kea, Hawaii. We also compare the results 
of this analysis with other sets of data taken at different wavelengths.
Abell 1835 is one of the most luminous clusters observed in the ROSAT 
catalogue and is well-known as a cooling core cluster.  
It is also known to contain two lensed sub-mm point sources, SMM J14009+0252 and SMM 
J140104+0252 (see e.g. \citet{Ivinson00, Zemcov07}).

The paper is organized as follows: Section 2 describes the observations with the Bolocam 
instrument; Section 3 describes the analysis 
pipeline developed for processing the data; Section 4 describes the modelling
of the data and the determination of the characteristic parameters of the model (as 
well as the errors in their values), and Section 5 discusses the results of the 
analysis of the Bolocam data and their combination with other literature results.

\section{Observations}

The data used in this work represents approximately 12.5 hours of 
observations taken over the course of five nights at the end of January/ 
beginning of February 2006. The observations were made with Bolocam operating at 
1.1 mm (275 GHz).

Bolocam consists of an array of 105 operational neutron-transmutation-doped 
(NTD) Germanium spiderweb bolometers, capable of observing at 1.1 mm or 
2.1 mm \citep{Glenn98, Haig04}. The system is cooled to 270 mK to allow the array to 
operate close to the photon background limit at 1.1 mm. The Bolocam array has a field
of view of approximately 8 arcmin, and the beam FWHM is $\sim$ 30{\it '' \/} at 
1.1 mm (and $\sim$ 60{\it '' \/} at 2.1 mm).

The detector is mounted at the Cassegrain focus of the Leighton telescope - the 
10.4 m CSO dish. The principal science targets for the instrument include 
star forming regions in the galaxy, blank-field surveys for dusty extragalactic
point sources; blank field SZ 
cluster surveys, and pointed observations of galaxy clusters.

Observations by Bolocam incorporate a number of different scan strategies. 
The majority of observations use raster scanning or lissajous scanning where the entire 
telescope is constantly in motion modulating the array position on the sky.

The data presented in this paper was taken using jiggle mapping. Jiggle mapping 
uses a combination of 'chopping' the secondary mirror during 
observations, i.e. switching the secondary from side-to-side, and 'nodding' the telescope, 
i.e. changing the position of 
the telescope during a scan. The beam moves from being 
'on-source' (imaging the region of the target and chopping away from it), to 'off-source' 
(imaging a nearby region of blank sky and chopping onto the target), then returns to being 
on-source once again (see Fig.~1).

\begin{figure}
 \includegraphics[width=3.3in]{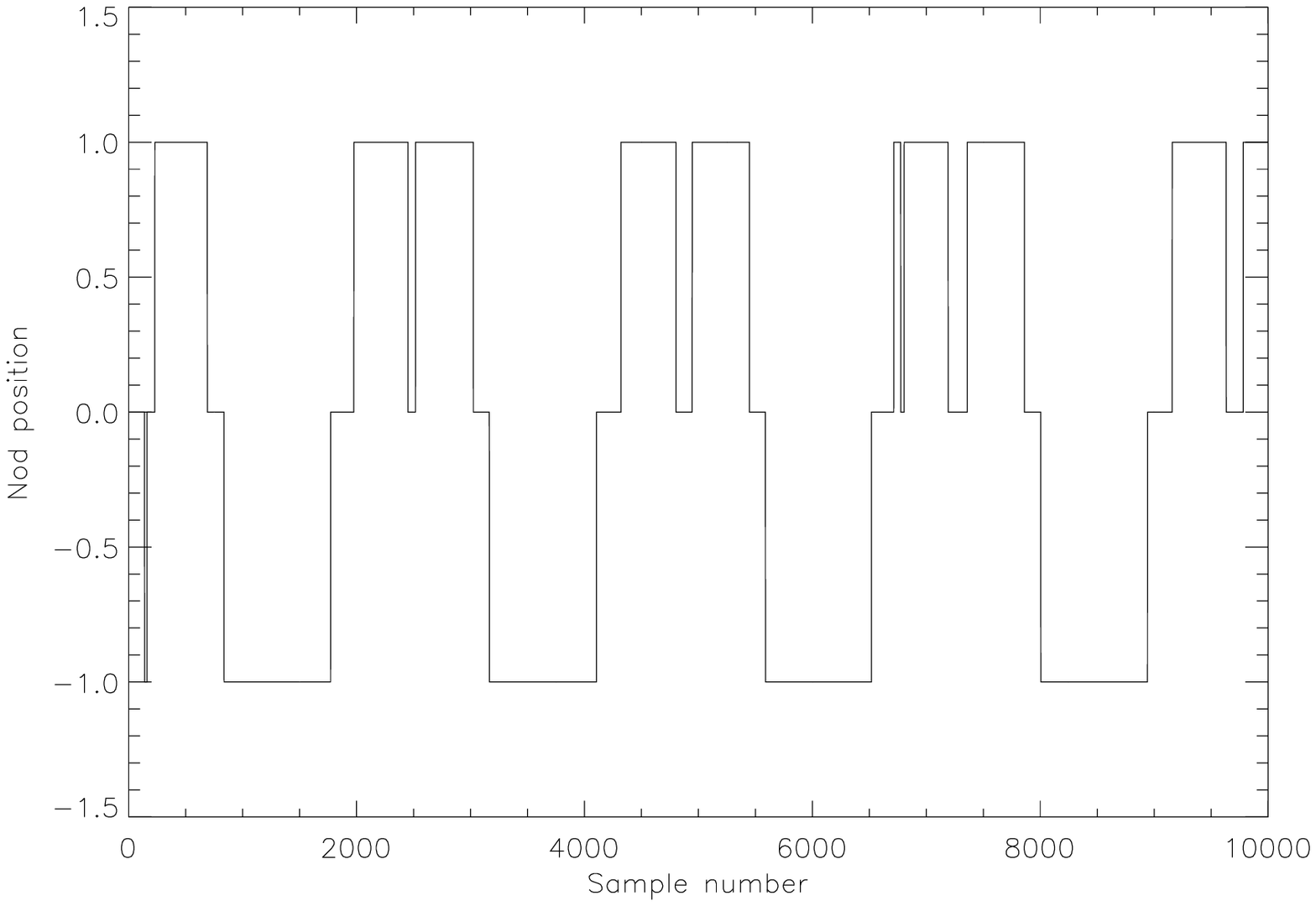}
 \includegraphics[width=3.3in]{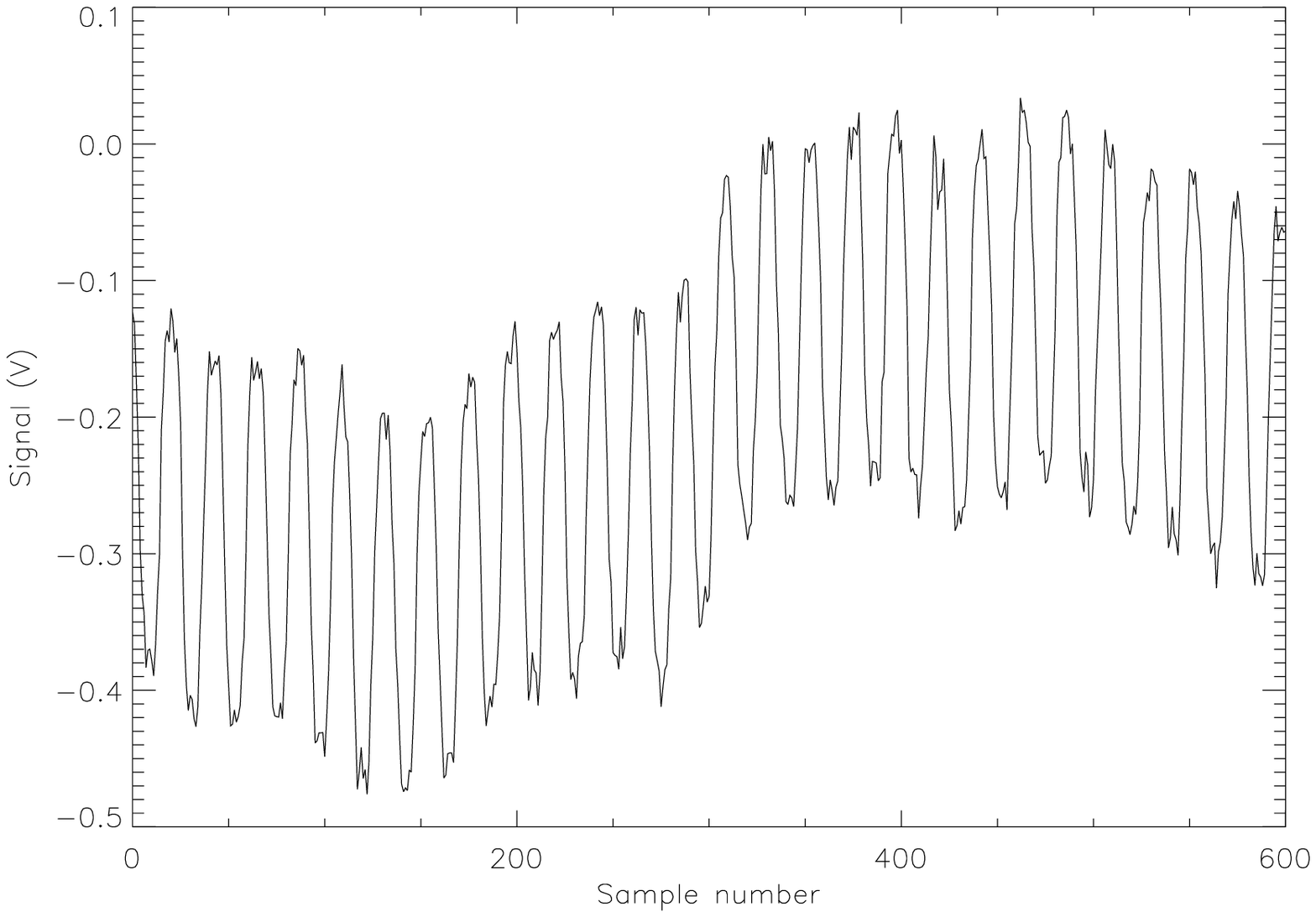}
 \caption{Nodding and acquisition pattern (top). A nod position of +1 implies the 
          telescope is on-source, whereas nod position -1 is when the telescope 
          is off-source. The pattern is a combination of acquisition and positional 
          data, so that where the pattern falls to zero this reflects where the 
          telescope is not acquiring data (usually while the dish is moving to a 
          different location on the sky). A sample of the raw data demonstrating 
          the chopping is also given (bottom).}
\end{figure}

Jiggle mapping is a means of removing sky background from data, and works by differencing 
signal from so-called 'target' and 'reference' beams (centered on the science target 
and a region of 'blank' sky, respectively). These beams are also referred to as 'on' 
and 'off' beams. While jiggle-mapping is an efficient method of subtracting sky-noise, 
care has to be taken to ensure that the chop 'throw' (i.e. the maximum angular dispacement 
of the secondary during the chop) is large enough that the off beam does not 
pick up signal from the source itself (subject to mechanical limitations). 
For this reason, jiggle-mapping is not always suitable for imaging 
extended sources, but is good for observing point sources.

The chop frequency for this data set was 2.25 Hz, giving a chop period of 
$\sim$ 0.44 s (compared to a sampling rate of 0.02 s). The chop throw was 
90'', while the displacement between on and off beams during the nods 
was 2.5'. The telescope typically remained in the 'on' or 'off' 
phase for 
periods of approximately 10 seconds.

The full set of data consisted of fifteen individual sets of observations of 
approximately 50 minutes each. The observations were separated by shorter 
observations of planets and other sources with known position and flux, which 
could be used for the calibration and pointing.

\section{Pipeline development}

Because jiggle-mapping is a relatively new strategy for Bolocam, there was 
initially no pipeline in place to analyse the raw data. An independent pipeline 
was written specifically for this observing mode.

In its original form, the time-stream data had been sliced (i.e. separated into
individual data-vectors representing the signal from individual bolometers during 
specific observations), but no other processing had taken place. The pipeline for jiggle data includes the following elements:

1. Initial cleaning to remove obvious sources of noise using average subtraction;

2. Identifying the nodding sequence;

3. Identifying phase differences between the signal from the chop control and the 
data (due, for example, to differences in timing between the telescope clock and 
those in the telescope control terminals);

4. Deconvolving the data for each nod position ('removing' the chopping);

5. Extracting the signal (differencing the deconvolved signal from the nod 
positions);

6. Calculating errors for the signals;

7. Determining pointing corrections for each map and hence reconstructing the 
directional information for each map;

8. Calibrating the maps;

9. Saving the deconvolved datasets,

10. Coadding maps from individual observations.

\subsection{Deconvolving}

We deconvolve the chopped data using a fit to the time stream signal monitoring the 
chopper position corrected for a phase difference between the chopper data and bolometer data. If 
the signal being detected by 
an individual bolometer is modulated by the chopping according to:

\begin{equation}
   \rmn{F}(t)=\rmn{F}_{\rmn{s}}\sin[\nu_{\rmn{c}}t+\phi_{\rmn{c}}]
\end{equation}
 where
 $\rmn{F}(t)$            is the time($t$)-varying flux received by the bolometer,
 $\rmn{F}_{\rmn{s}}$     is the flux of the source,
 $\nu_{\rmn{c}}$         is the chop frequency,
 $\phi_{\rmn{c}}$        is the chop phase, one can then multiply this by a fit and integrate according to:

\begin{equation}
   \rmn{S}(t)={\int\rmn{F}(t)\sin[\nu_{\rmn{f}}t+\phi_{\rmn{f}}]\rmn{d}t\over\int\sin^{2}[\nu_{\rmn{f}}t+\phi_{\rmn{f}}]\rmn{d}t}
\end{equation}
where
 $\nu_{\rmn{f}}$   is the fit frequency,
 $\phi_{\rmn{f}}$  is the fit phase.

With an accurate enough fit $(\nu_{\rmn{f}}\sim\nu_{\rmn{c}}$, 
$\phi_{\rmn{f}}\sim\phi_{\rmn{c}}$), therefore, $S(t)\rightarrow\rmn{F}_{\rmn{s}}$. The value of $\nu_{\rmn{f}}$ is measured from a fit to the chopper data, and the integration needs to be over a complete number of chopping cycles to avoid spurious signal being introduced into the maps.

\subsection{Pointing}

The individual detector pointing data was reconstructed in three stages. First a 
map was made of the deconvolved 
data. The directional information was taken from the time-streams of azimuth and 
elevation stored with the raw data (converted to ra and dec) and previously 
determined offsets (to convert the values stored at the telescope with the actual 
azimuth and elevation of the centre of the telescope beam) were added. The mean of 
these values were compared to the recorded position of the source to obtain a 
'mean' offset.

The directional information for the map was then reconstructed to include the mean 
offsets. The location of the centre of emission for the source in the map was 
determined using a simple $\chi^2$ fit of the data immediately around the 
cluster to a gaussian. This position was then compared once more with the source 
values to determine a 'fine' offset.

The directional information was then undated again to include both the mean and 
fine offsets.

\subsection{Calibration}

The calibration for Bolocam is determined using the resistances of the bolometers as
determined by
the dc level of the lock-in amplifier output signal, which is expected to be directly 
related to the flux calibration for a set of observations (see \cite{Laurent05}). 
If observations can 
be made of a number of sources of known flux at different levels of sky loading 
a plot can be made of the dc level 
against calibration factor. We fit a second-order polynomial to the measured calibration 
points to extend the calibration to the full range of observation conditions. In this way,
we can determine the correct calibration for an observation within an appropriate 
range of dc level.

These calibration curves are not expected to change significantly over time. However, 
given that the previous set of data was taken in May 2004, using a different scanning 
technique, we performed a full set of calibration observations roughly every 20 minutes 
during the science observations. These calibration sources were chosen to be close to 
the science observation target and included: 0420m014; 0923p392$\_$4cp39.25; 
1334$-$127, and 
3c371. The resulting calibration curve is given in Fig.~2.

The observations of the calibrators agree reasonably well with the May 2004 calibration 
curve. The point sources used in the plot were
all secondary sources and as a result it was not clear how precise or stable their
fluxes would be. Given this, the deviations 
between the old calibration curve and the sources were not considered to be great 
enough to warrant revising the calibration values and adopting a new functional form
for the calibration, and the May 2004 calibration was used throughout.

\begin{figure}
 \includegraphics[width=3.3in]{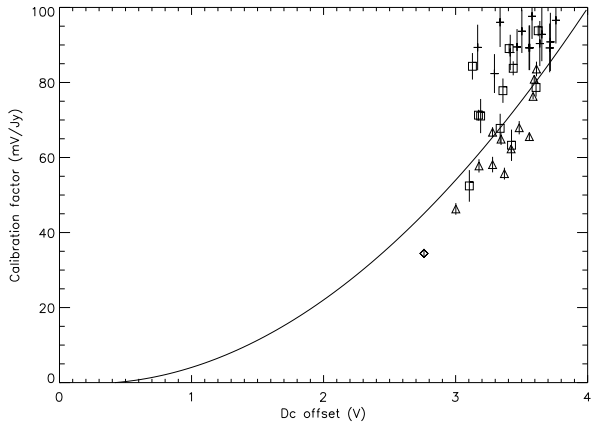}
 \caption{Calibration data. The May 2004 calibration curve is shown as the solid line.
          The error bars in this figure are statistical errors from the measured data. They do not, therefore, include intrinsic uncertainty in the source brightness. The data does not deviate substantially from the May 2004 
          curve. Legend: Vertical crosses: 0420m014; open triangles: 0923p392$\_$4cp39.25; 
          open diamonds: 1334-127, open squares: 3c371.}
\end{figure}

\section{Modelling and parameter estimation}

Once maps had been made of the individual observations, they were coadded to produce 
a single image of Abell 1835 which was then also convolved with a Gaussian PSF with 
FWHM $\sim$ 30.6{\it `` \/}, corresponding to the best-fitting Gaussian to the Bolocam beam 
(Fig.~3).

\begin{figure*}
 \includegraphics[width=3.4in]{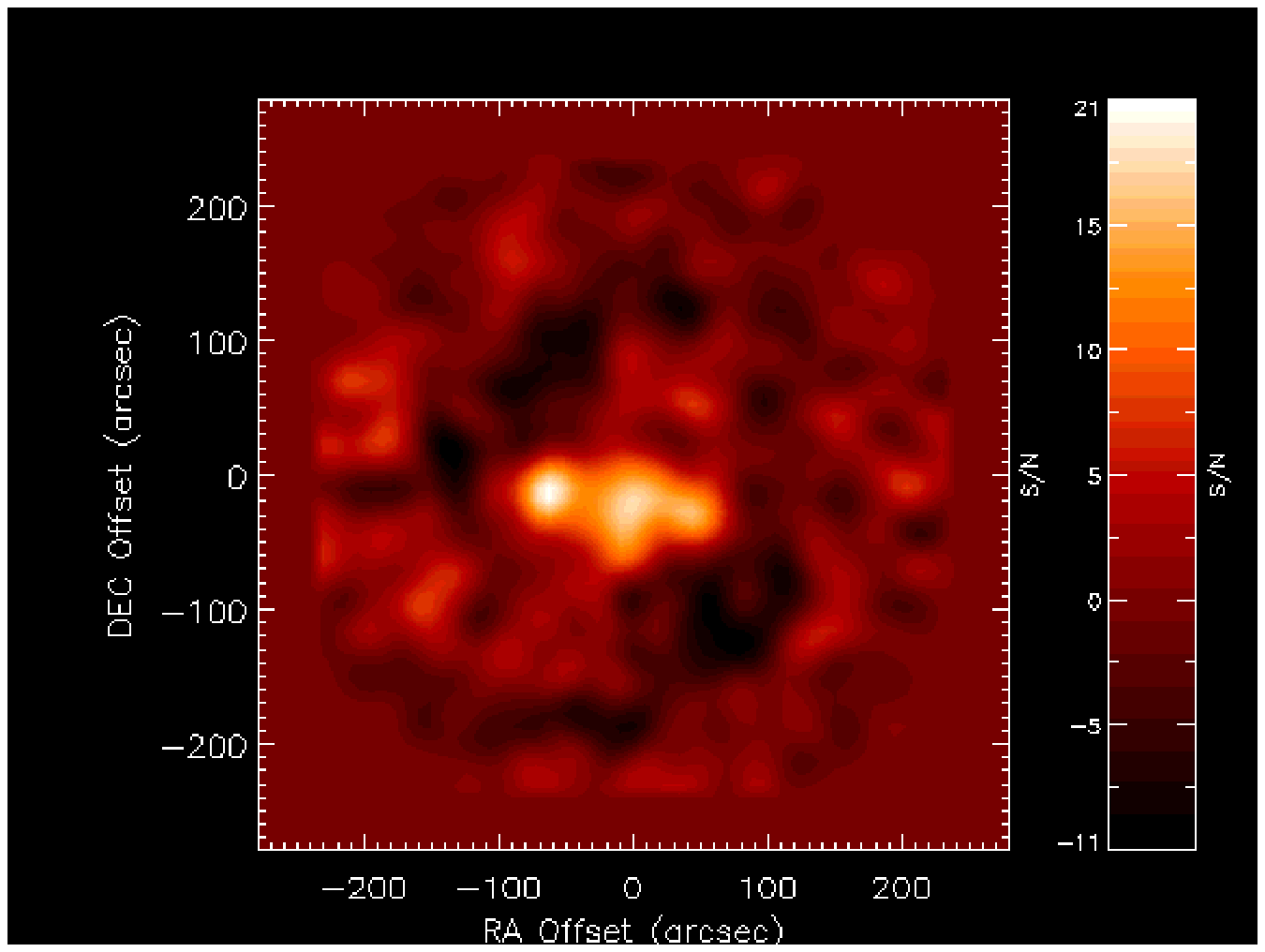}
 \includegraphics[width=3.5in]{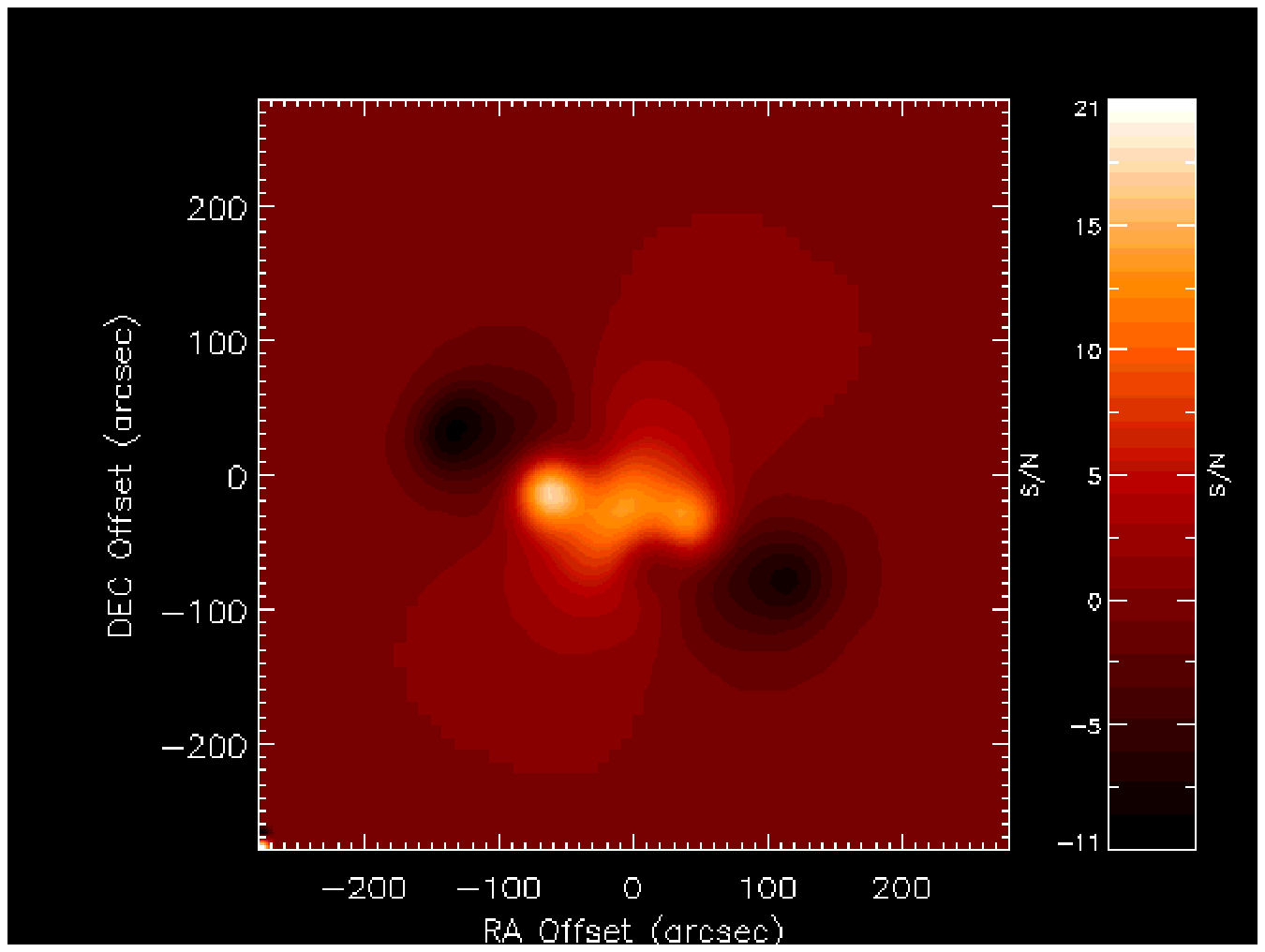}
 \includegraphics[width=3.5in]{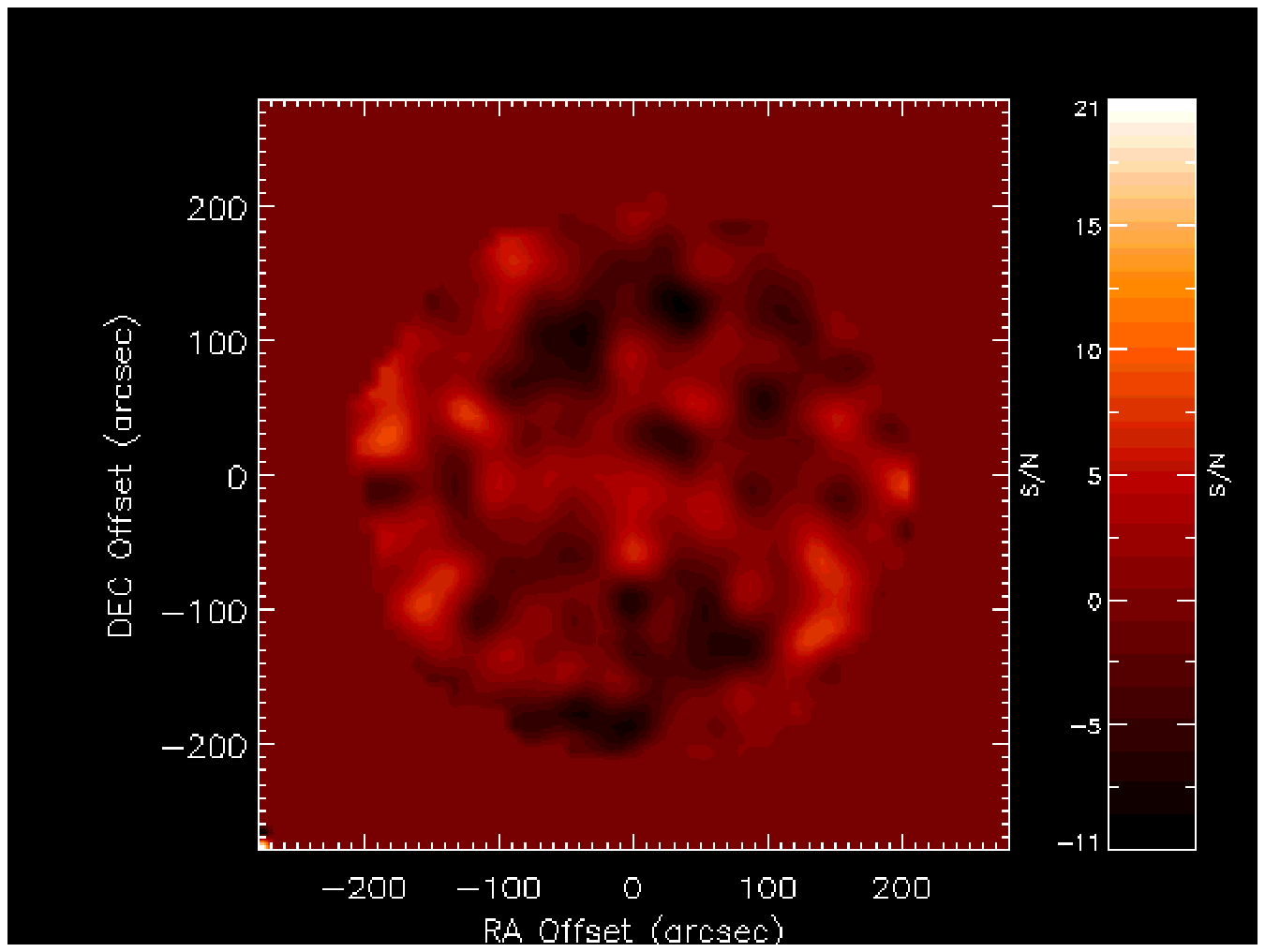}
 \caption{(Top left) S/N map of the raw data, convolved with the Bolocam beam; (top 
           right) 3-parameter fit ($\beta=0.69$, $\theta_{c}=33.6''$) S/N map convolved 
           with the Bolocam beam, (bottom) Map of residuals, formed by taking the 
           difference between the convolved model and data S/N maps.}
\end{figure*}

A model was produced to simulate the observations and derive estimates of the fluxes 
of the sources. Fig.~3 clearly shows the two point sources arranged on either side of 
the emission from the cluster (which is assumed to be entirely signal from the SZ). These point 
sources have been detected previously at 850 $\mu$m using the Submillimeter Common User 
Bolometer Array (SCUBA) on the James Clerk Maxwell Telescope (JCMT) \citep{Zemcov07} as well 
as by \citet{Ivinson00}. The characteristics of these point sources, reported by \citet{Zemcov07} 
are given in Table~1.

\begin{table*}
 \centering
 \begin{minipage}{140mm} 
  \begin{tabular}{@{}lcccc@{}}
  \hline
  Source id    & ra (hh:mm:ss) & dec (dd:mm:ss) & 850 $\mu$m Flux (mJy) & 450 $\mu$m Flux (mJy) \\
 \hline
 SMM J14011+0252 & 14:01:04.7 & +02:52:25 & 14.6$\pm$1.8 & 41.9$\pm$6.9 \\
 SMM J14009+0252 & 14:00:57.5 & +02:52:49 & 15.6$\pm$1.9 & 32.7$\pm$8.9 \\
\hline
\end{tabular}
\caption{Point source positions and fluxes reproduced from \citet{Ivinson00}} 
\end{minipage}
\end{table*}

The SZ emission from the galaxy cluster itself was modelled assuming that the signal was dominated 
by the thermal SZ and assuming an isothermal beta radial profile:

\begin{equation}
\mathcal{F}(\theta_{p})=\rmn\it{A}\left(1+\frac{\theta_{p}}{\theta_c}\right)^{\left(\frac{1}{2}-\frac{3 \beta}{2}\right)}
\end{equation}
where
 $\mathcal{F}(\theta_{p})$   is the flux profile of the cluster as a function of projected 
angle $\theta_{p}$,
 $\theta_{c}$ and $\beta$   are fitting parameters
 $\rmn\it{A}$               is a scaling parameter, defined by:

\begin{equation}
   \rmn{A}=\Delta\Omega\rmn{I}_{\tiny{0}}\rmn{y}_{0}\rmn{g(x)}\times10^{26}
\end{equation}
Here: 
  $\Delta\Omega$     is the solid angle of the pixels in the (model) map (the map was later 
convolved with the Bolocam beam such that the final flux measurements were given in flux/beam),
  $I_{0}=\frac{2h}{c^2}\left(\frac{k_{B}T_{CMB}}{h}\right)^3$,
is the blackbody emission of the CMB at the redshift of the cluster,
  $\rmn{y}_{0}$      is the so-called 'Compton parameter' at the centre of the cluster 
(the Compton parameter in general, $\rmn{y}$, can be characterised as:

\begin{equation}
   \rmn{y}=\int{\rmn{n}_{e}\sigma_{T}\frac{\rmn{k}_{B}\rmn{T}_{e}}{\rmn{m_{e}}\rmn{c}^2}}\rmn{dl}
\end{equation}
where
 $\rmn{n}_{e}$ and $\rmn{T}_{e}$ are, respectively, the electron density and electron 
temperature along the line of sight), and
  $\rmn{g(x)}$   is a function that depends on the dimensionless frequency at which 
the observations are being made, $\rmn{x}=\frac{\rmn{h}\nu}{\rmn{k}_{B}\rmn{T}_{\rmn{CMB}}}$, 
as:

\begin{equation}
   \rmn{g(x)}=x^4\frac{\exp{x}}{(\exp{x}-1)^2}(x\coth{(x/2)}-4)
\end{equation}

The temperature reported in the literature for Abell 1835 varies between approx. 
8 - 10 keV (e.g. \cite{Allen98, Peterson01, Katayama04}). At these temperatures, 
relativistic corrections to the SZ can be on the 
order of $\sim$ 10$\%$ and have to be taken into account during the analysis. 
We used the corrections 
published by \cite{Itoh98}, up to fifth order in $\theta$, where:

\begin{equation}
  \rmn{\theta}=\frac{\rmn{k}_{B}\rmn{T}_{e}}{\rmn{m}_{e}\rmn{c}^2}
\end{equation}

An ideal sky model was produced including extended emission from the cluster 
and emission from the point sources at positions given by \cite{Zemcov07}. The 
free parameters of the model were the cluster geometry ($\theta_{c}$ and $\beta$); 
the comptonization ($y_{0}$), and the fluxes of the two point sources $\mathcal{F}_{1}$ 
and 
$\mathcal{F}_{2}$ (corresponding to SMM J0140104+0252 and SMM J14009+0252, 
respectively). For each set of parameters, we produced a high resolution map with 
pixel size of 1". This map was then convolved with the Bolocam 
beam to produce a 'real-sky' map of the cluster at the resolution of the telescope. 
This in turn was then 'observed' by using the data ra and dec values to reproduce the 
effect of the jiggling, thereby producing a simulation of what we expect to observe 
given the assumed values of flux (for the point sources) and $\rmn{y}_{0}$. Finally, 
the chopped map was rebinned to the same pixel size as the raw data maps. By comparing 
the maps corresponding to different values 
of the parameters with the data, we construct a multi-parameter likelihood space, which 
could be searched to extract best fitting models.

Values for these parameters were determined by searching for the set 
that minimized the $\chi^2$ value between the model map and the data map.

Generally, SZ observations are not able to constrain the parameters $\theta_{c}$ and $\beta$, 
since the resolution of such experiments is not accurate enough to constrain their values 
to a better degree of accuracy than equivalent X-ray (or other wavelength) measurements. 
Recently, however, computer simulations have suggested that using X-ray derived profiles 
for fitting SZ observations can lead to bias in the derived values of $\rmn{y}$ 
\citep{Hallman07}. Furthermore, the values of the profile parameters quoted in the literature 
can vary considerably (see, e.g. values reported in \citet{Jia04, Peterson01, Schmidt01, Zemcov07}).

In this work, therefore, a range of approaches have been adopted. Fits to $\rmn{y}$, 
$\mathcal{F}_{1}$ and $\mathcal{F}_{2}$ were made using two different sets of values for 
$\theta_{c}$ and $\beta$ obtained using Chandra X-ray data and interferometric SZ data from 
the Owens Valley Radio Observatory (OVRO) and the Berkeley Illinois Maryland Association 
(BIMA), taken from\citet{LaRoque06}.

The cores of galaxy clusters can exhibit 
sharp peaks in emission due to the presence of cooling flows, and for this reason a 
single isothermal beta model fit to all the X-ray data in a set of observations is not 
always suitable. \citet{LaRoque06} use two different strategies to deal with this. In the first 
the X-ray data set is restricted to exclude the region within 100 kpc of the X-ray centre, 
and the profile parameters are then obtained by a joint fit to this X-ray data and the 
entire set of SZ data. It is possible to use the entire SZ dataset since SZ emission has a 
weaker dependence upon electron density than X-ray observations, and the SZ profile is 
not expected to be significantly affected by the presence of cooling flows. (Indeed in 
general, the SZ profile is less sensitive to the detailed physics in the cores of galaxy 
clusters \citep{Motl05}.) In the second approach they employ a double-beta model which has 
been used to fit all the X-ray data in observations that include emission from cooling 
flows (see e.g. \citet{Mohr99}). \citet{LaRoque06} also fit to the SZ data only (although this fit 
is really only to $\theta_{c}$ since the value of $\beta$ is fixed to the value obtained 
for the isothermal beta X-ray/ SZ fit).

For the purpose of this analysis the 
results from the isothermal beta fit to the restricted X-ray data and SZ data (hereafter 
LaRoque X-ray+SZ) and the fit to the SZ data (hereafter LaRoque SZ only) were used. The 
values of $\theta$ and $\beta$ they obtain for these fits are, for the combined X-ray and 
SZ data: $\beta=0.69$, $\theta_{c}=33.6''$, and for the SZ-only data: $\beta=0.70$, 
$\theta_{c}=50.1''$.  

It was decided, therefore, to fit the data from the Bolocam observations with five 
combinations of profile parameters: $\beta=0.69$, $\theta_{c}=33.6''$; $\beta=0.69$, 
$\theta_{c}=50.1''$; $\beta=0.69$; $\theta=33.6''$, and $\theta=50.1''$, where the 
other parameters are fitted for in each situation.

The goodness of fit for a particular set of parameters was evaluated using a $\chi^{2}$ 
test. The value for $\chi^{2}$ for a given model was defined simply as:

\begin{equation}
\chi^2=\displaystyle\sum_{i}{\frac{(\rmn{data}_{i}-\rmn{model}_{i})^2}{\sigma_{i}}}
\end{equation}

where $\rmn{data}_{i}$ denotes the signal in the i'th pixel of the data map, $\rmn{model}_{i}$ represents 
the signal in the i'th pixel of the model map, and $\sigma_{i}$ is the noise in the i'th 
pixel of the noise map. The best fitting parameters for a given analysis run were, therefore, 
those that minimized this value of $\chi^{2}$.

For the cases in which there were four free parameters to be searched through, the 
individual searches took a long time to process. The general method of locating a minimum 
was to sample the parameter space between 'reasonable' limits, locate a minimum and examine 
variation in the $\chi^{2}$ values for each parameter while keeping the other parameters 
fixed. This variation was in general smooth and reasonably quadratic, so that the first 
derivatives of the variation was typically linear. This could easily be fit and an estimate 
made of where the first derivative was zero. The limits of the search were then reset 
to centre on the new approximation for the minimum and the process repeated until the new 
estimates of the minima were similar (within some tolerance) to the previous ones. This 
approach allowed the minimum values to be located within a reasonable amount of time.

Errors on the values of the fit parameters were found by deriving the inverse of the Fisher 
matrix for the observations and evaluating its diagonal elements. For parameters with a 
gaussian distribution, the Fisher matrix ${\it F}_{ij}$ is given by:

\begin{equation}
 {\it F}_{ij}=\frac{1}{2}\left(\frac{\partial^2\chi^2}{\partial{p_i}\partial{p_j}}\right)
\end{equation}
where 
   $p_i$ represents the vector of parameters being fit,
   and the second-order differential is evaluated at the value of $p$ at which the $\chi^2$ 
value is minimized.

This kind of analysis relies on there being no covariance between different pixels in the data 
maps. The level of covariance is expected to be low due to the chopping, and because the only 
cleaning of the data involves an average subtraction. This should introduce less pixel to pixel 
covariance than other cleaning methods. Jacknife maps were also made to check these assumptions, 
and were found to show no evidence for significant levels of covariance. 

The components of ${\it F}_{ij}$ were evaluated using a finite difference scheme and arrays of 
$\chi^2$ values around the minimum values. The $\chi^{2}$ values were obtained by varying the fit 
parameters as well as the position of the central emission. The results of this analysis are 
shown in Table~2.

\begin{table*}
\label{ResultsTable}
 \begin{minipage}{140mm}
 \begin{tabular}{l|cccccc}
 $\beta_{model}$ & 0.69 & 0.69 & 0.69 & -- & -- & 0.69 (p) \\
 $\left(\theta_{c}\right)_{model}$ ($''$) & 33.6 & 50.1 & -- & 33.6 & 50.1 & 33.6 (p) \\
 \hline
 $\mathcal{F}_{1}$ (mJy) & 6.48$\pm{2.00}$ & 6.22$\pm{2.08}$ & 7.16$\pm{1.88}$ & 7.09$\pm{1.89}$ & 7.17$\pm{1.84}$ & 6.48$\pm{2.00}$ \\
 $\mathcal{F}_{2}$ (mJy) & 11.32$\pm{1.92}$ &  10.90$\pm{1.98}$ & 11.98$\pm{1.79}$ & 12.38$\pm{1.82}$ & 12.18$\pm{1.77}$ & 11.32$\pm{1.92}$ \\
 $\rmn{y}_{0}\times{10^{-4}}$ & (4.68$\pm{0.48}$) & (4.43$\pm{0.49}$) & (6.52$\pm{0.64}$) & (4.85$\pm{0.47}$) & (4.19$\pm{0.40}$) & (4.34$\pm0.52$) \\
 $\beta$ & 0.69 & 0.69 & 0.69 & $1.20\pm{0.2}$ & $1.60\pm{0.2}$ & 0.69 \\
 $\theta_{\rmn{c}}$ ($''$) & 33.6 & 50.1 & $14.7^{+2.4}_{-2.2}$ & 33.6 & 50.1 & 33.6 \\
 ra offset ($''$) & 1.8 & 1.9 & 2.1 & 2.1 & 2.0 & 2.0 \\
 dec offset ($''$) & 3.6 & 3.9 & 2.6 & 2.8 & 2.5 & 3.4 \\
 $\chi^2$ & 3066 & 3079 & 3058 & 3056 & 3054 & 3062 \\
 \hline
 \end{tabular}
 \caption{Parameter estimates (1 mm Bolocam observations) and $\chi^{2}$ values; Number 
           of degrees of freedom = 3013. (p) denotes that the model included a 1.8 mJy point source
           at the cluster centre.}
 \end{minipage}
\end{table*}

\section{Results}

The results of Table~2 shows that the 4-parameter fit with $\theta_{c}=50.1''$ gives the best ($\chi^{2}$)
 value. The best 
fit values of the central Compton parameter as well as the fluxes of the point sources are, therefore, 
found to be: $y_{0}=(4.19\pm0.48)\times10^{-4}$; 7.17$\pm$1.84 mJy, and 12.18$\pm$1.77 mJy respectively. The 
errors on the values of $y_{0}$ in Table~2 include statistical errors as well as errors due to pointing and 
uncertainty in the model parameters. They do not include systematic errors (e.g. due to dust contamination, 
kinetic SZ, etc.). The systematic uncertainties are discussed in greater detail below.

The value of $\beta$ for our best fit model is substantially different from values quoted by other groups, 
who use X-ray 
data with much better resolution to calculate their profile parameters. It is common practice to treat these 
measurements as more reliable in determining cluster profile parameters (although see \cite{Hallman07}). Furthermore, 
since a comparison of our data to that of other 
groups was planned, the cluster model for which intensity measurements were being compared had to be the same. 
For these reasons, it was decided that, in forming the comparison with data at other frequencies, the results 
for the more 'canonical' values the profile parameters should be used. Of the two 
fits with more typical profile parameters, the $\beta=0.69$, $\theta_{c}=33.6''$ fit has the lower $\chi^2$ 
value, and is the model used for the spectral fitting.

\section*{Additional sources in the field}

In order to reliably simulate a series of chopped observations, all sources in the field need to be 
accounted for. The $\beta=0.69$, $\theta_{c}=33.6''$ fit model is also shown in Fig.~3, convolved 
with the Bolocam beam. The final map in Fig.~3 displays the difference 
between the model map and the raw data, convolved with the Bolocam beam to emphasize any differences 
between them. The difference map appears to indicate that the cluster signal 
and the point sources have been removed effectively and that, therefore, the model is a reasonable 
simulation of the data. The signal at the edge of the field is mostly spurious. It could be suggested
that there is a $\sim$ 3-$\sigma$ detection (in the raw map) of a point source just below centre of the 
image field, but 
there are no obvious listed sources located at the position of the candidate detection, and it appears 
that most likely to be a spurious feature of the noise.

Using noise maps calculated in the analysis, it was possible to form a histogram of the difference between 
the S/N data map and the simulated map (see Fig.~4, where the data is plotted on logarithmic axes to determine 
whether there was significant deviation between the residuals and the fit toward the limits of the data). The 
Gaussian form of the histogram (quadratic in Fig.~4) is 
consistent with the residual signal being dominated by noise, rather than other point sources. The histogram 
was fit with a Gaussian, 
and the best fit has a standard deviation close to 1.0, indicating that the noise had been correctly estimated. 
There may be some suggestion of deviation in the high S/N region between the histogram and the fit, but this is 
attributed to regions of lower coverage in the outer sections of the map, and not believed to represent a 
real excess of sources.

\begin{figure}
 \includegraphics[width=3.3in]{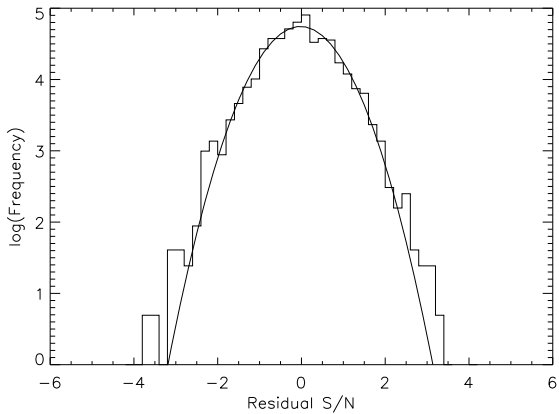}
 \caption{Signal to noise histogram of the residual map obtained by subtracting the best fit model from the 
          data, plotted on logarithmic scales. The Gaussian best fit to the histogram (the solid line in the log 
          plot) has $\sigma=1.01$}
\end{figure}

\section*{Dust Contribution}

At longer wavelengths the contribution from dust emission is expected to
become more important and is a source of confusion. The literature does not
provide an entirely consistent view of whether or not Abell 1835 has
significant dust emission. While the cluster's brightest cluster galaxy (BCG) exhibits
significant CO emission which is believed to trace dust, it would be
expected that the dust emission would be extremely bright at 450 $\mu$m.
There are a limited number of results in the literature that report emission from
Abell 1835 at 450 $\mu$m. \cite{Ivinson00} and \cite{Edge99} report strong emission, consistent with a significant
dust contribution, whereas the 450 $\mu$m result given in \cite{Zemcov07} of -2 $\pm$13 mJy is consistent with
there being no emission from the cluster at this wavelength.

Assuming a 'worst case' scenario , however, in which the emission reported by \cite{Ivinson00} at 450 and 850 $\mu$m are 
assumed to be correct, and all the signal at these wavelengths is attributable to the cD galaxy, a model can be formed 
of the dust emission using a greybody spectrum, assuming a dust 
temperature $\sim$ 30 K and an emissivity which varies as $\nu^{1.5}$.
The flux of the cD galaxy at 1.1mm using this model is expected to be $\sim1.8\pm0.5$ mJy. This value is consistent with the SED produced for Abell 1835's BCG reported by \cite{Egami06}. Introducing a point source 
into the cluster 
model at the X-ray centre and then marginalizing again over values of $y_{0}$, we obtain a best fit value for 
$y_{0}$ of $(4.34\pm0.52)\times10^{-4}$, with a $\chi^{2}$ value of $\sim$ 3062.

\begin{figure}
\label{SZspectrum}
\includegraphics[width=3.5in]{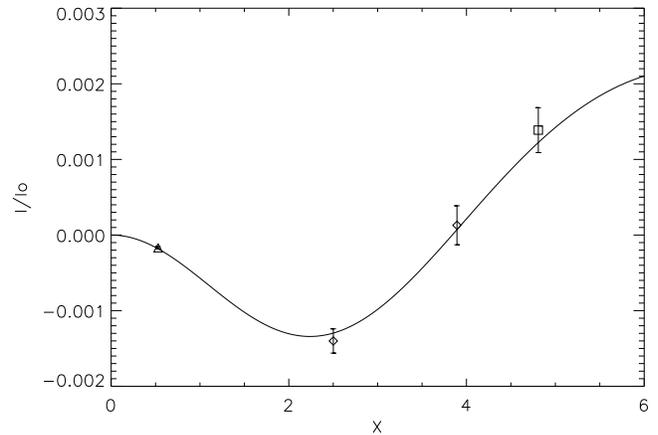}
\caption{The SZ spectrum of Abell 1835.
The 30 GHz results (open triangle, $x \sim 0.5$) are taken from.LaRoque, et al. (2006).
The points at 142 and 221~GHz (open diamonds) are from Mauskopf, et al. (2000),
while the open square represents the results
reported here. All points have been renormalised to the central intensity of the SZ effect for a cluster model with $\theta_0 = 33.6''$
and $\beta = 0.69$. The line is the best fit SZ spectrum with $y_0 = 3.60 \times 10^{-4}$ and $v_z = -226$~km/s.}
\end{figure}

\begin{table*}
\label{error_budget}
 \begin{minipage}{70mm}
  \begin{tabular}{@{}l|c@{}}
  \hline
  Systematic errors  & Error (as $\%$ of result)\\
 \hline
 Calibration & 11.7 \\
 Kinematic effect & 9.0 \\
 CMB confusion & 1.0 \\
 Dusty galaxies & 5.4 \\
 Dust emission & 7.2 \\
 Effective temperature & $^{+1.5}_{-2.6}$ \\
 Uncertainties in & 0.6 \\
 King model parameters & \\
 Others (clumping, etc) & 2 \\
 \\
 \hline
 Pointing and model & 0.52 mJy \\
 error (incl. point & \\
 source & \\
 Total systematic & 0.69 mJy \\
 Total error & 0.87 mJy \\
 \hline
\end{tabular}
\caption{Error budget for Abell 1835 observations}
\end{minipage}
\end{table*}

\section*{Other sources of error}

Along with the dust contribution, it was important to account for other sources of error in the result. These are 
discussed individually below:

\subsection*{Calibration}

We estimated the calibration error from the noise-weighted dispersion of the measured point fluxes relative to the May, 2004 model. This gives a calibration error of $\sim$ 10.6$\%$, 
which, when combined with a 5$\%$ error due to uncertainties in the Mars model that all the Bolocam fluxes are dependent 
on gave an overall calibration error of $\sim$ 11.7$\%$.

\subsection*{Kinematic effect}

The kinematic SZ effect (see below) is due to the bulk motion of a cluster. In general, the kinematic effect 
is much less significant than the thermal effect that has been discussed so far. The bulk motion of clusters in a 
concordance cosmology predicts an rms peculiar velocity of around 300~km/s, but \cite{Mauskopf00} obtain an estimate 
of the velocity of Abell 1835 of around 500~km/s. A velocity this high at 273 GHz produces a kinematic effect 
that is approxiamtely 9$\%$ of the thermal effect.

\subsection*{Confusion sources}

The main sources of confusion in observations of the SZ effect include the CMB background and dusty galaxies. Typical CMB temperature anisotropy signals based on the current concordance model spectrum was simulated and 'observed' using our scan strategy. These produced an rms equivalent to a 1.5$\%$ error in the measured cluster signal. Dusty galaxies at 1.1 mm contribute an rms signal of approximately 0.5 mJy (e.g. \cite{Blain98}), which corresponds to a 5.4$\%$ error in the measured cluster signal.

\subsection*{Physical model uncertainties}

Other sources of error related to the parameters of the physical model of Abell 1835 itself. Uncertainties in the 
electron gas temperature of 10 - 15 $\%$ correspond to an error in the estimate of $y_{0}$ of up to $\sim$ 2.5$\%$. 
\cite{LaRoque06} reports errors on the model parameters $\theta_{c}$ and $\beta$ of $\pm$1.0'' n and $\sim$ 0.01, 
respectively. When these variations are introduced into the model to see what effect they have on the value of $y_{0}$ 
obtained by the fit, their effect is found 
to be small, and represent an uncertainty on the level of 0.6 $\%$ on the final result. Other 
potential sources of error include (e.g.) clumping in the cluster gas, but the SZ effect is relatively insensitive to 
the detailed physics of cluster gas, these effects were estimated as contributing no more than $\sim$ 2$\%$ to the final 
result.

A breakdown of the error budget for the observations presented here is given in Table~3.

\section*{$y_{0}$ estimates}

The value of $y_{0}$ given by the 4 parameter fit means that the observations detect the cluster with a significance of 
10.5$\sigma$. This is one of the highest significance detections of a cluster in the positive region of the SZ spectrum 
to date.
A plot of the SZ spectrum of Abell 1835 based upon these values is given in Fig.~5. These values have were 
fitted with a spectrum that included kinetic and thermal effects (non-relativistic as well as relativistic). The 
kinetic effect follows the form:

\begin{equation}
\label{Kinetic}
I_{kinetic}=-\tau_{e}\beta h(x)
\end{equation}

where $\tau$ is the optical depth of the cluster gas, given by $y_{0}\left(\frac{mc^{2}}{k_{B}T_{e}})\right)$, 
$\beta\equiv\frac{v_{z}}{c}$ (where $v_{z}$ is  the peculiar velocity of the galaxy cluster), and 
$h(x)=x^4\frac{\exp{x}}{(\exp{x}-1)^2}$. The kinetic effect dominates the SZ spectrum around the null point 
($\nu\sim217$ GHz, x $\sim$ 1.9), but makes only a small contribution to the spectrum at frequencies above or below 
this.

The free parameters of the fit to the SZ spectrum were, therefore, $y_{0}$ (which characterizes the thermal SZ), and 
$v_{z}$ (which characterizes the kinetic effect). The values of these parameters that optimized the fit were found to be: 
$y_{0}=(3.60\pm0.24)\times10^{-4}$, and $v_{z}=-226\pm275$~km/s. This is consistent with previous measurements of 
$v_{z}$ for Abell 1835 \cite{Benson04, Mauskopf00} and represents the most sensitive test for peculiar velocity 
in an individual galaxy cluster using the SZ effect to date.

\section{Conclusions}

The observation and analysis of 1.1 mm Bolocam jiggle-map data of Abell 1835, including 
details of the process of converting the raw data to maps of the cluster, have been 
discussed. A parameter search has been used to determine the best fit values of the central 
Compton parameter for the cluster and the fluxes for the point sources SMM J140104+0252 and 
SMM J14009+0252. These values are found to be $y_{0}=(4.68\pm0.48\pm0.82)\times 10^{-4}$, 
6.5$\pm{2.0}$ mJy and 11.3$\pm{1.9}$ mJy, respectively with a statistical signal-to-noise of 
10.5, 3.3 and 6.0 respectively. If the model is refined further to include a point source, 
representing dust emission from the cD galaxy in Abell 1835, the Compton parameter for the 
cluster is found to be $y_{0}=(4.34\pm0.52\pm0.69)\times 10^{-4}$ The value for $y_{0}$ was 
compared to other literature results 
and found to agree well. The SZ spectrum for Abell 1835, based upon these values for $y_{0}$ 
was fit with the full SZ form, including both thermal and kinetic effects and relativistic 
corrections up to 7th order. The values of $y_{0}$ and $v_z$ which optimized this fit were 
found to be $(3.60\pm0.24)\times 10^{-4}$ and $-226\pm 275$~km/s, which are both in 
agreement with previous literature results. It is also concluded that, in order to better 
evaluate the contamination from dust and accurately characterize the 
spectrum of point sources in the same field as SZ clusters more results need 
to be obtained at shorter wavelengths in order to obtain more a detailed spectral 
energy distribution.

\section*{Acknowledgments}

This work was supported by NSF grants AST-9980846 and AST-0206158. We wish to acknowledge M. Zemcov for providing access to data and for useful discussion, which improved the content of this paper significantly. We would also like to recognize and acknowledge the cultural role and reverence that the summit of Mauna Kea has within the Hawaiian community. We are fortunate and privileged to be able to conduct observations from this mountain.

\bsp

\label{lastpage}


\begin{thebibliography}{99}

\bibitem[\protect\citeauthoryear{Albrecht et al.}{2006}]{Albrecht06} Albrecht A., Bernstein G., Cahn R. et al., 2006, preprint(astro-ph/0609591)
\bibitem[\protect\citeauthoryear{Allen \& Fabian}{1998}]{Allen98} Allen S. W., Fabian A. C., 
MNRAS, 297, L57, 1998 
\bibitem[\protect\citeauthoryear{Battistelli et al.}{2003}]{Battistelli03} Battistelli E. S., De Petris M., Lamagna L. et al., 2003, preprint(astro-ph/0303587)
\bibitem[\protect\citeauthoryear{Benson et al.}{2004}]{Benson04} Benson B. A., Church S. E., Ade P. A. R., Bock J. J., Ganga K. M., Henson C. N., Thompson K. L., 2004, ApJ, 617, 829
\bibitem[\protect\citeauthoryear{Bhattacharya \& Kosowsky}{2007}]{Bhattacharya07} Bhattacharya S., Kosowsky A., 2007, preprint (astro-ph/0712.0034)
\bibitem[\protect\citeauthoryear{Birkinshaw, Hughes \& Arnaud}{1991}]{Birkinshaw91} Birkinshaw M., Hughes J. P., Arnaud K. A., 1991, ApJ, 376, 466
\bibitem[\protect\citeauthoryear{Birkinshaw}{1999}]{Birkinshaw99} Birkinshaw M., 1999, Phys. Rep., 310, 97
\bibitem[\protect\citeauthoryear{Blain}{1998}]{Blain98} Blain A. W., 1998, MNRAS, 297, 502
\bibitem[\protect\citeauthoryear{Carlstrom, Holder \& Reese}{2002}]{Carlstrom02} Carlstrom J. E., Holder G. P., Reese E. D., 2002, ARA\&A, 40, 643
\bibitem[\protect\citeauthoryear{DeDeo, Spergel \& Trac}{2005}]{DeDeo05} DeDeo S., Spergel D. N., Trac H., 2005, preprint(astro-ph/0511060)
\bibitem[\protect\citeauthoryear{Diego et al.}{2002}]{Diego02} Diego J. M., Martinez-Gonzalez E., Sanz J. L., Benitez, Silk J., 2002, MNRAS, 331, 556
\bibitem[\protect\citeauthoryear{Edge et al.}{1999}]{Edge99} Edge A. C., Ivinson R. J., Smail I., Blain A. W., Kneib J. -P., 1999, MNRAS, 306, 599
\bibitem[\protect\citeauthoryear{Egami et al.}{2006}]{Egami06} Egami E., Misselt K. A., Rieke G. H. et al., 2006, ApJ, 647, 922
\bibitem[\protect\citeauthoryear{Glenn et al.}{1998}]{Glenn98} Glenn J., Bock J. J., Chattopadhyay G. et al., 1998 in, Proc. SPIE, 3357, 326
\bibitem[\protect\citeauthoryear{Grainge}{1996}]{Grainge96} Grainge K., 1996, PhD thesis, 
Cambridge University
\bibitem[\protect\citeauthoryear{Grego et al.}{2001}]{Grego01} Grego L., Carlstrom J. E., Reese E. D., Holder G. P., Holzapfel W. L., Joy M. K., Mohr J. Patel S., 2001, ApJ, 552, 2
\bibitem[\protect\citeauthoryear{Haig et al.}{2004}]{Haig04} Haig D. J., Ade P. A. R., Aguirre J. E. et al.,
2004, Proc. SPIE, 5498, 78
\bibitem[\protect\citeauthoryear{Hallman et al.}{2007}]{Hallman07} Hallman E. J., Burns J. O., Motl P. M., Norman M. L., 2007, preprint (astro-ph/0705.0531)
\bibitem[\protect\citeauthoryear{High et al.}{2010}]{High10} High F. W., Stalder B., Song J. et al., 2010, preprint(astro-ph/1003.0005)
\bibitem[\protect\citeauthoryear{Hincks et al.}{2009}]{Hincks09} Hincks A. D., Acquaviva V., Ade P. A. R. et al., 2009, preprint(astro-ph/0907.0461)
\bibitem[\protect\citeauthoryear{Holzapfel et al.}{1997}]{Holzapfel97} Holzapfel W. L., Arnaud M., Ade P. A. R. et al., 1997, ApJ, 480, 449
\bibitem[\protect\citeauthoryear{Itoh, Kohyama \& Nozawa}{1998}]{Itoh98} Itoh, N., 
Kohyama, Y., Nozawa, S. 1998, ApJ, 502, 7
\bibitem[\protect\citeauthoryear{Ivison et al.}{2000}]{Ivinson00} Ivison R. J., Smail I., Barger A. J., Kneib J.-P., Blain A. W., Owen F. N., Kerr T. H., Cowie L. L., 2000, MNRAS, 315, 209
\bibitem[\protect\citeauthoryear{Jia et al.}{2004}]{Jia04} Jia S. M., Chen Y., Lu F. J., Chen L., Xiang F., 2004, A\&A, 423, 65
\bibitem[\protect\citeauthoryear{Jones}{1995}]{Jones95} Jones M., ApL Comm., 1995, 
32, 347
\bibitem[\protect\citeauthoryear{Katayama \& Hayashida}{2004}]{Katayama04} Katayama H., 
Hayashida K., Advances in Space Research Vol. 34, 12, 2519, 2004
\bibitem[\protect\citeauthoryear{LaRoque et al.}{2006}]{LaRoque06} LaRoque S. J., Bonamente M., Carlstrom J. E., Joy M. K., Nagai D., Reese E. D., Dawson K. S., 2006, ApJ, 652, 917
\bibitem[\protect\citeauthoryear{Laurent et al.}{2005}]{Laurent05} Laurent G. T., Aguirre J. E., Glenn J. et al., 2005, ApJ, 623, 742
\bibitem[\protect\citeauthoryear{Mauskopf et al.}{2000}]{Mauskopf00} Mauskopf P. D., Ade P. A. R., Allen S. W. et al., 2000, ApJ, 538, 505
\bibitem[\protect\citeauthoryear{Mohr, Mathiesen \& Evrard}{1999}]{Mohr99} Mohr 
J. J., Mathiesen B., Evrard A. E., 1999, ApJ, 517, 627
\bibitem[\protect\citeauthoryear{Motl et al.}{2005}]{Motl05} Motl P. M., Hallman E. J., Burns J. O., Norman M. L., 2005, ApJ, 623, L63
\bibitem[\protect\citeauthoryear{Peterson et al,}{2001}]{Peterson01} Peterson J. R., Paerels F. B. S., Kaastra J. S. et al., 2001, A\&A, 365, L104
\bibitem[\protect\citeauthoryear{Plagge et al.}{2009}]{Plagge09} Plagge T., Benson B. A., Ade P. A. R., Aird K. A., Bleem L. E., 2009, preprint(astro-ph/0911.2444)
\bibitem[\protect\citeauthoryear{Reese}{2003}]{Reese03} Reese E. D., 2003 in ed. Freedman W. L., Carnegie Observatories Astrophysics Series, Vol. 2, Cambridge University Press
\bibitem[\protect\citeauthoryear{Rephaeli, Sadeh \& Shimon}{2006}]{Rephaeli06} Rephaeli 
Y., Sadeh S., Shimon M., 2006, preprint (astro-ph/0511626)
\bibitem[\protect\citeauthoryear{Schmidt, Allen \& Fabian}{2001}]{Schmidt01} Schmidt 
R. W., Allen S. W., Fabian A. C., 2001, MNRAS, 327, 1057 
\bibitem[\protect\citeauthoryear{Sunyeav \& Zel'dovich}{1970}]{Sunyaev70} Sunyaev R. A., 
Zel'dovich Ya. B., 1970, Ap\&SS, 7, 3
\bibitem[\protect\citeauthoryear{Tsuboi et al.}{1998}]{Tsuboi98} Tsuboi M., Miyazaki A., Kasuga T., Matsuo H., Kuno N., 1998, PASJ, 50, 169
\bibitem[\protect\citeauthoryear{Udomprasert et al.}{2004}]{Udomprasert04} Udomprasert P. S., Mason B. S., Readhead A. C. S., Pearson T. J., 2004, pre-print (astro-ph/0408005)
\bibitem[\protect\citeauthoryear{Weller, Bettye \& Kneissl}{2002}]{Weller02} Weller J., Battye R. A., Kneissl R., 2002,  Phys. Rev. Lett., 88, 231301
\bibitem[\protect\citeauthoryear{Zemcov et al.}{Zemcov et al.}{2007}]{Zemcov07} Zemcov M., Borys C., Halpern M., Mauskopf P. D., Scott D., 2007, MNRAS, 376, 1073

\end{thebibliography}
\end{document}